\documentstyle[prl,aps,twocolumn,epsfig]{revtex}
\begin{document}
\draft
\twocolumn[
\hsize\textwidth\columnwidth\hsize\csname@twocolumnfalse\endcsname

\title{Drastic facilitation of the onset of global chaos in a periodically 
       driven Hamiltonian system due to an extremum in the dependence of 
       eigenfrequency on energy}

\author{S.M. Soskin$^{1,2}$, O.M.~Yevtushenko$^{3}$,
and R. Mannella$^{2,4}$
}

\address{
$^1$Institute of Semiconductor Physics,
National Academy of Sciences of Ukraine, Kiev, Ukraine\\
$^2$Department of Physics, Lancaster University, Lancaster
LA1 4YB, UK\\
$^3$ ICTP, Trieste, Italy\\
$^4$Dipartimento di Fisica, Universit\`{a} di Pisa and INFM UdR Pisa,
56100 Pisa, Italy
}

\date{\today} \maketitle \widetext

\begin{abstract}
The Chirikov resonance-overlap criterion predicts the onset of global
chaos if nonlinear resonances overlap in energy, which is conventionally
assumed to require a non-small magnitude of perturbation. We show that,
for a time-periodic perturbation, the onset of global chaos may occur at
unusually {\it small} magnitudes of perturbation if the unperturbed system
possesses more than one separatrix. The relevant scenario is the
combination of the overlap in the phase space between resonances of the
same order and their overlap in energy with chaotic layers associated with
separatrices of the unperturbed system. One of the most important
manifestations of this effect is a drastic increase of the energy range
involved into the unbounded chaotic transport in spatially periodic system
driven by a rather {\it weak} time-periodic force, provided the driving
frequency approaches the extremal eigenfrequency or its harmonics. We
develop the asymptotic theory and verify it in simulations.
\end{abstract}

\pacs{05.45.Ac, 05.45.Pq, 75.70.cn}
]
\narrowtext

A weak perturbation of a
Hamiltonian system
causes the onset of chaotic layers in the phase space around separatrices of the
unperturbed system and/or separatrices surrounding nonlinear
resonances
generated by the perturbation itself
\cite{Chirikov:79,lichtenberg_lieberman,Zaslavsky:1991}:
the system may be transported along the layer random-like.
Chaotic transport plays an important role in many physical
phenomena \cite{Zaslavsky:1991}.
If the perturbation is weak enough, then the layers are thin and such chaos is called
{\it local} \cite{Chirikov:79,lichtenberg_lieberman,Zaslavsky:1991}. As the
magnitude of the perturbation increases,
the widths of
the layers grow and the layers corresponding to adjacent separatrices (either
to those of the unperturbed system, or to separatrices surrounding
different nonlinear resonances)
reconnect at
some, typically non-small, critical value of the magnitude, which conventionally marks
the onset of {\it global} chaos \cite{Chirikov:79,lichtenberg_lieberman,Zaslavsky:1991}
i.e. chaos in a large range of the phase space which increases
with a further increase of the magnitude of perturbation. Such
scenario often correlates with the overlap in {\it energy} between
neighbouring resonances calculated {\it independently} in the resonant
approximations of the corresponding orders: this constitutes the
celebrated empirical Chirikov resonance-overlap criterion
\cite{Chirikov:79,lichtenberg_lieberman,Zaslavsky:1991}.

But the Chirikov criterion may fail in time-periodically perturbed {\it zero-dispersion}
(ZD) systems \cite{soskin94,soskin97,pr} (cf. also studies of related maps
\cite{Howard:84,Howard:95}) i.e. systems in which the frequency of
eigenoscillation possesses an extremum (typically, a local maximum or
minimum) as a function of its energy. In such systems, there are
typically two resonances of one and the same order \cite{more}, and their overlap in energy does not
result in the onset of global chaos \cite{soskin97,pr,Howard:84,Howard:95}. Even their
overlap in phase space results typically, instead of the onset of global
chaos, only in reconnection of the
thin chaotic layers associated with the resonances while, as the
amplitude of the time-periodic perturbation grows further, the layers
{\it separate} again (with a different topology \cite{soskin97,pr,Howard:84,Howard:95})
despite the {\it growth} of the width of the overall relevant
range of energy.

The major idea of the {\it present} work is to {\it limit} the
energy range relevant to the overlap of resonances of the same
order by chaotic layers relating either to resonances of a different order \cite{same} or to
separatrices of the unperturbed system (these scenarios immediately suggest the corresponding substitute
for the Chirikov criterion in a ZD system in the range of extremum). The onset
of global chaos occurs respectively in a broader energy range or at {\it unusually small}
magnitudes of the perturbation. The former case will be considered
elsewhere while we present below details of the latter case. It
relates to many important systems and the above effect is manifested in a very strong
manner: {\it all} systems with more
than one separatrices possess the ZD property \cite{soskin97,pr}
and global chaos may involve the whole energy range between the
separatrices.

We use
as an example a potential system with a periodic
potential possessing two different-height barriers in a period
(Fig. 1(a)):

\begin{eqnarray}
&&
H_0(p,q)=\frac{p^2}{2}+U(q), \quad\quad U(q)=\frac{(\Phi-\sin(q))^2}{2},
\\
&&
\Phi={\rm const}<1.
\nonumber
\end{eqnarray}

The model (1) may describe e.g. a 2D electron gas in a
magnetic field spatially periodic in one of dimensions of the gas \cite{oleg98}.
The interest to this system arose due to technological advances of
the last decade allowing to manufacture high-quality magnetic
superlattices \cite{Oleg12,Oleg10} leading to a variety of interesting
behaviours of charge carriers in semiconductors
(see \cite{oleg98,Oleg12,Oleg10,Shmidt:93,shepelyansky} and references
therein). The same model describes a pendulum on a hinge spinning
about its vertical axis \cite{andronov}.

The separatrices of the Hamiltonian system (1) in the $p-q$ plane and the dependence of the
frequency $\omega$ of its
oscillation, often called {\it eigenfrequency}, on energy $E\equiv H_0(p,q)$ are shown in Figs. 1(b) and 1(c)
respectively. As seen from the latter, $\omega(E)$ is close to
the extreme eigenfrequency $\omega_m\equiv\omega(E_m)$ in the major part of the
interval $[E_b^{(1)},E_b^{(2)}]$ while
sharply decreasing to zero as $E$ approaches $E_b^{(1)}$ or
$E_b^{(2)}$. Such features of $\omega(E)$ are {\it typical} for
systems with two or more separatrices.

Let the time-periodic
perturbation be added:

\begin{eqnarray}
&&
\dot{q} = \partial H/\partial p, \quad\quad \dot{p} = -\partial
H/\partial q,
\\
&&
H(p,q)=H_0(p,q)- h q\cos (\omega_f t).
\nonumber
\end{eqnarray}

Let us first consider the {\it conventional} evolution of chaos in the system (2)-(1) as $h$ grows
while $\omega_f$ remains fixed at an arbitrarily chosen value beyond
the immediate vicinity of $\omega_m$ and its harmonics.
It is illustrated by
Fig.\ 2. At small $h$, there are two thin chaotic layers, around
the inner and outer separatrices of the unperturbed system.
Unbounded chaotic transport takes place only in the
outer chaotic layer i.e.\ in a {\it narrow} energy range. As $h$
grows, so also do the layers until, at some
critical value $h_{gc} \equiv h_{gc} (\omega_f)$, they
merge. This event may be considered as the onset of global chaos: the
whole range of energies between the barriers levels then becomes
involved in unbounded chaotic transport.
Note that
the states $\{I^{(l)} \}\equiv \{p=0,q=\pi/2 +2\pi l \}$ and
$\{O^{(l)} \}\equiv\{p=0,q=-\pi/2 +2\pi l\}$ (where $l=0,\pm 1,
\pm 2, ...$) in the stroboscopic (for instants $n2\pi/\omega_f$
with $n=0,1,2,...$) Poincar\'{e} section are associated respectively with the
inner and outer saddles of the unperturbed system, and
necessarily belong to the inner and outer chaotic layers
respectively. Thus, the necessary and sufficient condition for
global chaos to arise in the system may be formulated as the possibility
for the system placed initially in the state $\{I^{(0)} \} $
to pass beyond the neighbouring \lq\lq outer'' states, $\{O^{(0)} \}
$ and $\{O^{(1)} \} $, i.e.\ for $q(t\gg 2\pi/\omega_f)$ to become
smaller than $-\pi/2$ or larger than $3\pi/2$.

A diagram in the $h-\omega_f$ plane,  based
on the above criterion, is shown in Fig.\ 3. The lower boundary of the
shaded area represents the function $h_{gc} (\omega_f)$. It
has deep cusp-like local minima ({\it spikes}) at frequencies
$\omega_f=\omega_s^{(n)}$ that are slightly less than the odd
multiples of $\omega_m$,

\begin{equation}
\omega_s^{(n)} \approx\omega_m(2n-1), \quad\quad n=1,2,...
\end{equation}

\noindent The deepest minimum occurs at
$\omega_s^{(1)}\approx\omega_m$: $h_{gc} (\omega_ s^{(1)})$ is
approximately 40 times smaller than in the neighbouring pronounced local
maximum of $h_{gc} (\omega_f)$ at $\omega_f\approx 1$. As $n$ increases,
an
$n$th minimum becomes less deep.

The origin of the spikes
becomes obvious from the analysis of the evolution of the
Poincar\'{e} section as $h$ grows while $\omega_f \approx \omega_
s^{(1)}$. For $h=0.001$, one can see in Fig.\ 4(a) four chaotic
trajectories: those associated with the inner and outer
separatrices of the unperturbed system
\cite{Chirikov:79,lichtenberg_lieberman,Zaslavsky:1991} are coloured
green and blue respectively, while the trajectories
associated with the two nonlinear resonances of the 1st order
\cite{Chirikov:79,lichtenberg_lieberman,Zaslavsky:1991}
are indicated by
red and cyan (the corresponding attractors are
indicated respectively by crosses of the same colours). Examples of
non-chaotic (often called KAM
\cite{Chirikov:79,lichtenberg_lieberman,Zaslavsky:1991})
trajectories separating the chaotic ones are shown in brown. As
$h$ increases to $h=0.003$ (Fig.\ 4(b)), the blue and red chaotic
trajectories mix: the resulting chaotic layer is shown in blue. As
$h$ increases further (see Fig.\ 4(c), where $h=0.00475$), the latter layer merges
with
the layer associated with
the cyan chaotic trajectory (the resulting layer is
shown by blue) and, finally, as $h$ increases slightly more
(see Fig.\ 4(d), where $h=0.0055$), the latter layer merges with
the layer associated with the green
trajectory \cite{footnote2}, thereby marking the onset of global chaos as
defined above.


For most of the range $[E_b^{(1)},E_b^{(2)}]$, $\omega(E)$
is close to $\omega_m$ \cite{footnote3} (cf. Fig. 1(c)), so that
the overlap between the resonances of the {\it same} order may be described
within the resonance approximation. Given that it
plays the main role in the above scenario,
the function $h_{gc}(\omega_f)$
may also be evaluated
in the ranges of
the spikes
using the resonance approximation. The explicit
asymptotic (for $\Phi\rightarrow 0$) formulae for the minima
themselves,
in the lowest order of the parameters of smallness, are as
follows:

\begin{eqnarray}
\omega_s^{(n)}=\frac{(2n-1)\pi}{2\ln(4{\rm e}/\Phi)},
\quad\quad
n=1,2,...,
\\
 \Phi^2\ll 1,\quad\quad
\{10\ln(4{\rm e}/\Phi)\}^{-1}\ll 1
\nonumber
\end{eqnarray}

\noindent
(note that $\omega_m\approx \pi/\{2\ln(8/\Phi)\}$, so that
(4) does not contradict (3)),

\begin{eqnarray}
&&
h_{gc}(\omega_s^{(n)})= \frac{(2n-1)c\Phi}{2\sqrt{2}\ln(4{\rm e}/\Phi)},
\quad
c\approx 0.179, \ n=1,2,..., \cr
&& \Phi^2\ll 1, \quad \frac{2n-1}{\ln(4{\rm e}/\Phi)}\ll 1, \\
&& \frac{\ln\{10\ln(4{\rm e}/\Phi)/(2n-1)
\}}{10\ln(4{\rm e}/\Phi)/(2n-1)}\ll 1,
\nonumber
\end{eqnarray}

\noindent
where $c$ is the root of the equation

\begin{equation}
\ln
\left(
\frac{1+\chi(c)}{1-\chi(c)}
\right)
-2\chi(c)=c,
\quad \chi(c)\equiv \sqrt{1-4{\rm e}^{c-2}}.
\end{equation}

Eqs. (4)-(6) have been derived within the following self-contained
scheme. Let $\omega_f$ be close to $n\omega_m$ so that, for most
of the energy interval $[E_b^{(1)},E_b^{(2)}]$ except in
the close vicinity of barriers levels, the $n$th harmonic of
eigenoscillation is nearly resonant to the driving. Then a slow
dynamics of action $I$ and angle $\psi$ \cite{Landau:76} may be
described by the following auxiliary Hamiltonian (cf.
\cite{Chirikov:79,lichtenberg_lieberman,Zaslavsky:1991,soskin94,soskin97,pr,Howard:84,Howard:95}):

\begin{equation}
\tilde{H}(I,\tilde{\psi})=\int_{I(E_m)}^{I}dI\;
(n\omega-\omega_f)\;-\;
nhq_n\cos(\tilde{\psi})\;,
\end{equation}

\noindent
where $I\equiv I(E) =(2\pi)^{-1}\oint dqp$ is action (note also that
$dE/dI=\omega$ \cite{Landau:76}) and $\tilde{\psi}\equiv n\psi - \omega_f t$
is slow angle 
\cite{Chirikov:79,lichtenberg_lieberman,Zaslavsky:1991,soskin94,soskin97,pr,Howard:84,Howard:95}
while $q_n\equiv q_n(I)$ is the modulus of the $n$th Fourier
harmonic of coordinate (see the exact definition in
\cite{soskin97,pr}).

For a given value of $h$ and $\omega_f$, there are typically two
stable stationary points in a $2\pi$ band of the $I-\tilde{\psi}$ plane,
often called
{\it resonances} of the $n$th order. Each resonance is surrounded by a
{\it separatrix} which includes one (per a $2\pi$ band) unstable stationary point, called saddle.
If both saddles possess the same $\tilde{H}$
the
separatrices reconnect each other. This condition is satisfied along some
line $\omega_f(h)$ in the $h-\omega_f$ plane.
We need to find the point $(\omega_s, h_s)$ on this line such that
the lowest along the reconnected separatrices energy $E_l$
coincides with the upper energy boundary $E_{cl}^{(1)}$ of the
chaotic layer around the lower barrier level $E_b^{(1)}$ (one can
prove that the chaotic layer around the upper barrier level is
then necessarily overlapped by the resonance separatrices: cf.
Fig. 4(c)). It is known \cite{Zaslavsky:1991} that, if
$\omega_f\stackrel{<}{\sim}\omega_0$ where $\omega_0$ is the
eigenfrequency at the bottom of the wells, then
$\delta\equiv E_{cl}^{(1)}-E_b^{(1)}\sim h\omega_f/\omega_0$.
One can prove that, if $\Phi\rightarrow 0$, then the values of
$\omega_s$ and $h_s$ are not sensitive to the exact value of
$\delta$ within this range. Thus, the values of $\omega_s$ and
$h_s$ can be obtained using only the {\it resonant approximation}.

In the unperturbed case ($h=0$), the equations of motion (2)-(1) can
be solved explicitly, and $\omega(E)$ can be found then
explicitly too. As $\Phi\rightarrow 0$, it reduces to

\begin{eqnarray}
&& \omega(E)=\frac{\pi}{\ln
\left(
\frac{64}{(\Phi-\Delta E)(\Phi+\Delta E)}
\right)},
\quad\quad \Delta E\equiv E-\frac{1}{2},
\\
&& \Phi \ll 1.
\nonumber
\end{eqnarray}

\noindent
Substituting Eq. (8) for $\omega$ into Eq. (7) for $\tilde{H}$ and
assuming that
$q_{2n-1}\stackrel{\Phi\rightarrow 0}{\longrightarrow}\sqrt{2}/(2n-1)$
in the relevant energy range
(note that $q_{2n}=0$ due to the symmetry of the potential), one
can derive Eqs. (4)-(6) as the solution of the system of two algebraic
equations obtained from the condition of the equality of $\tilde{H}$
in the saddles and from the condition
that  $E_l-E_b^{(1)}\sim h\omega_f/\omega_0$,
keeping in the solution only those terms which have the lowest order in the
parameters of smallness. The solution confirms the
original assumptions that $\omega\approx \omega_f/(2n-1)$ and
$q_{2n-1}\approx \sqrt{2}/(2n-1)$
in the relevant range of energies, which
proves that the theory is self-contained. Note also that, to
logarithmic accuracy, the asymptotic formulas (4)-(6) are valid
for an arbitrary
symmetric potential (whose period is normalized to $2\pi$) and
they may be rather easily generalized for an {\it arbitrary} Hamiltonian
system with two separatrices close to each other.

The values of $\omega_s^{(n)}$ obtained from simulations for
$\Phi=0.2$ (see Fig.\ 3) nicely agree with Eq. (4). As concerns
$h_{gc}(\omega_s^{(n)})$, the
value $\Phi=0.2$ is too large for Eq.\ (5) to be accurate but,
even so, Eq.\ (5) provides a very good estimate for
$h_{gc}(\omega_s^{(1)})$ and the correct order of magnitude for $h_{gc}(\omega_s^{(2)})$ (see Fig.\
3). Calculations within the same scheme but using exact
formulas for $\omega$ and $q_{2n-1}$ and
solving the system of the algebraic equations
numerically yield the values
$(\omega_s^{(1)}=0.4\pm 0.004, \; h_{gc}(\omega_s^{(1)})=
0.005\pm 0.001)$ and $(\omega_s^{(2)}=1.243\pm 0.01, \; h_{gc}(\omega_s^{(2)})=
0.025\pm 0.008)$, nicely agreeing with the simulations.

It is worth mentioning two generalizations:

1.
For {\it non-symmetric} $U(q)$, $q_{2n}\neq 0$ and, therefore,
pronounced spikes in $h_{gc}(\omega_f)$ exist at $\omega_f\approx 2n\omega_m$
too.

2. If the time-periodic driving is {\it multiplicative},
then the resonances become {\it parametric} (cf.\
\cite{Landau:76}). Parametric resonances are more complicated and
much less studied than nonlinear resonance. But, still, the main
mechanism for the onset of global chaos remains the same, and
simulations show that $h_{gc}(\omega_f)$
do possess deep spikes. At the same time,
the frequencies of the main spikes are twice as large
as those of the corresponding
spikes
in the case of the additive driving: this is so because the
characteristic frequencies of parametric resonance are typically
doubled as compared with the nonlinear resonance (cf.\
\cite{Landau:76}).

Finally, we suggest two examples of physical applications: (i) a
jump-wise increase of the {\it dc
conductivity}
occurs due to
the jump-wise increase of the range of energies involved in the
unbounded chaotic transport of electric charge carriers in a magnetic
superlattice (cf. \cite{oleg98});
(ii) a significant decrease of the {\it activation energy}
for noise-induced multi-barrier escape in the presence of periodic driving
is associated with the
noise-free transport from the lower barrier to beyond the upper barrier
(cf.\ \cite{iric,pr}).

It is worth noting some analogies between our work and the work
\cite{percolation} describing the so called stochastic percolation
in 2D Hamiltonian systems where the merging of internal and external
chaotic zones was also relevant. However, both the models and the
underlying mechanisms are very different (there is no zero-dispersion situation in
\cite{percolation}; rather the two-dimensionality is inherently important); besides, in
\cite{percolation}, there was no explicit quantitative
description analogous to our Eqs. (4)-(6).

The work was supported by INTAS Grants 97-574 and 00-00867.
We are grateful to J. Howard, P. McClintock, J. Meiss and K. Richter for discussions.

\newpage

\begin{figure}
\begin{center}
\unitlength1cm
\begin{picture}(6,5)
   \epsfig{file=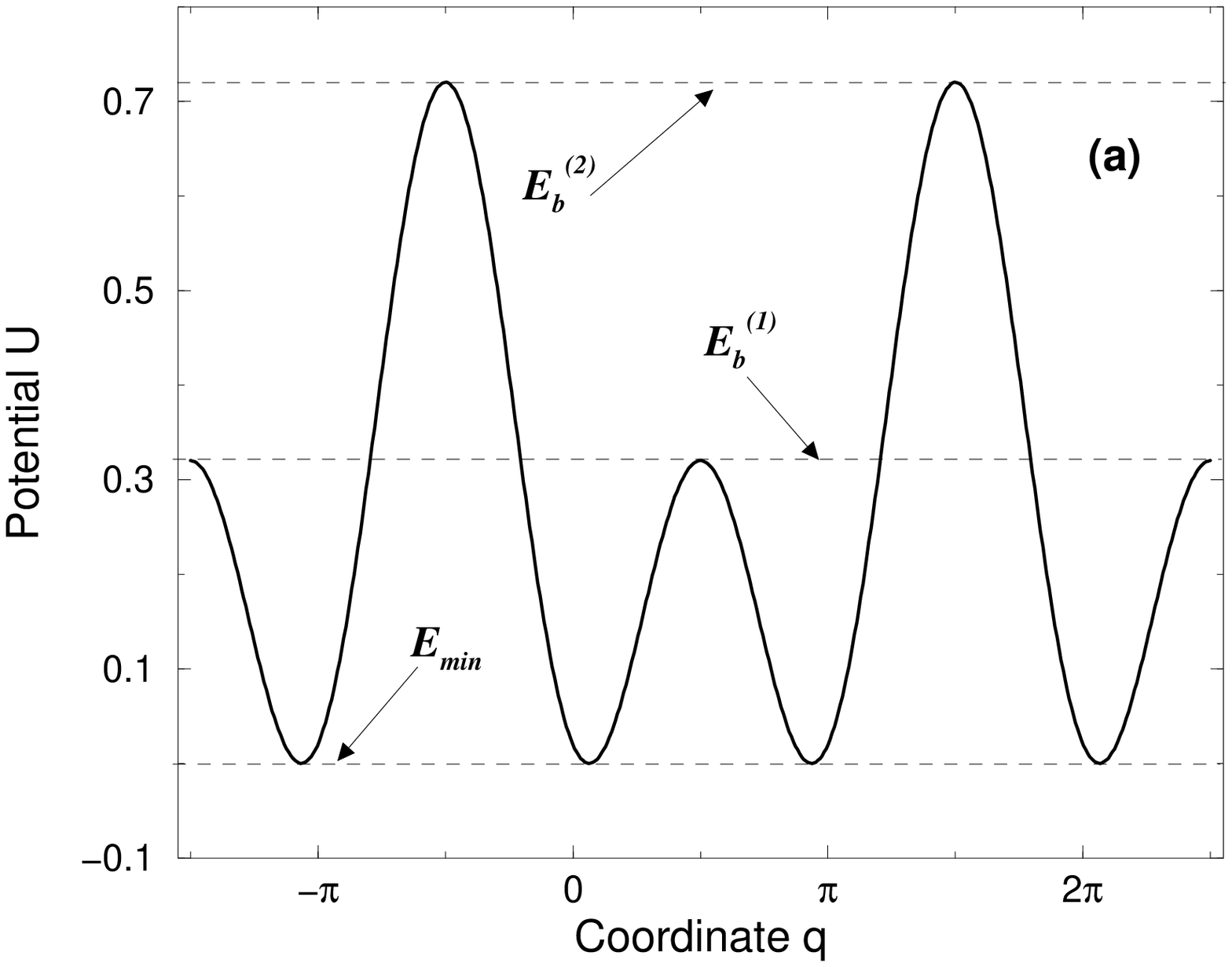,angle=0,width=6cm}
\end{picture}
\begin{picture}(6,4)
   \epsfig{file=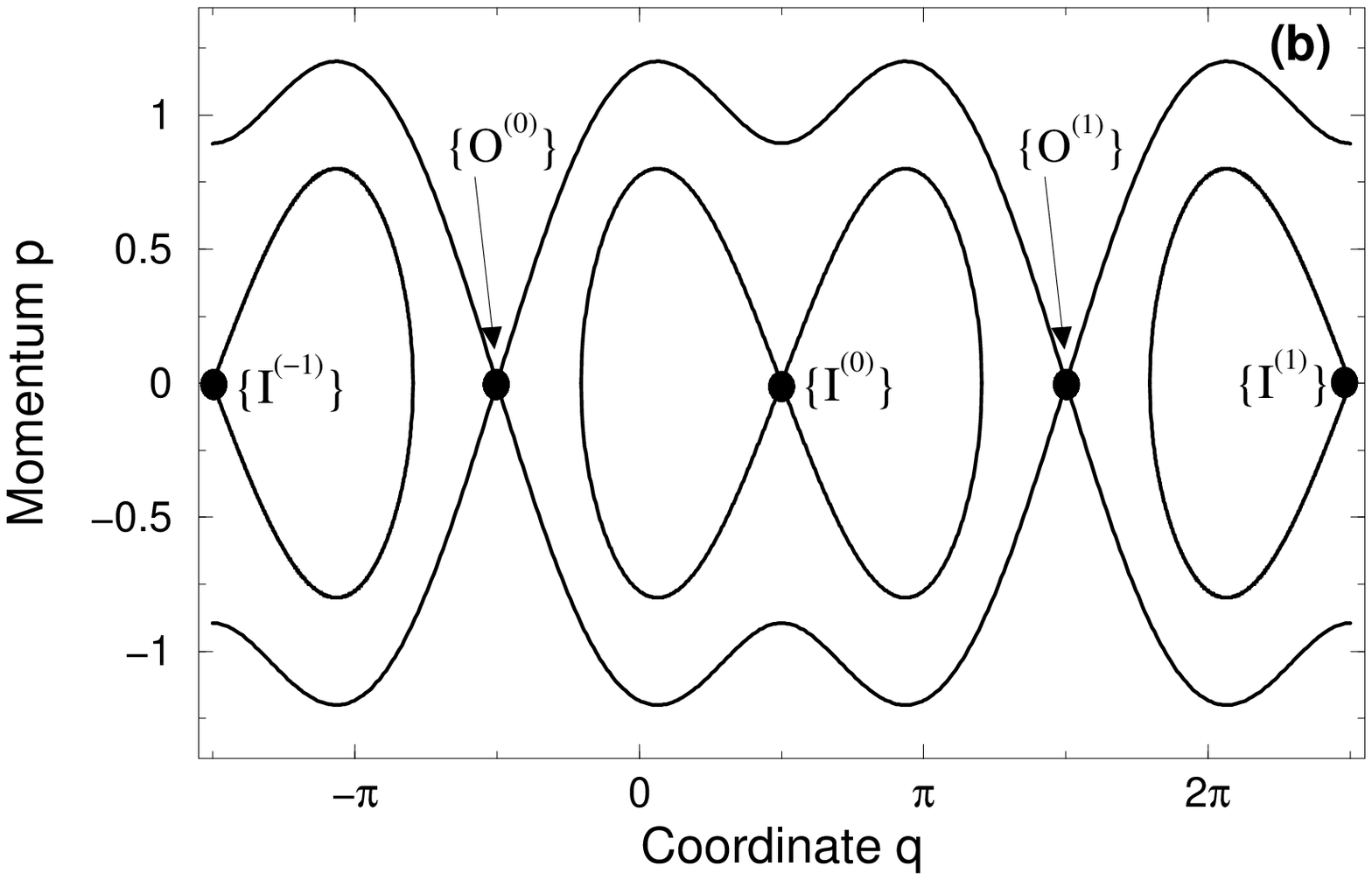,angle=0,width=6cm}
\end{picture}
\begin{picture}(6.25,4,5)
   \epsfig{file=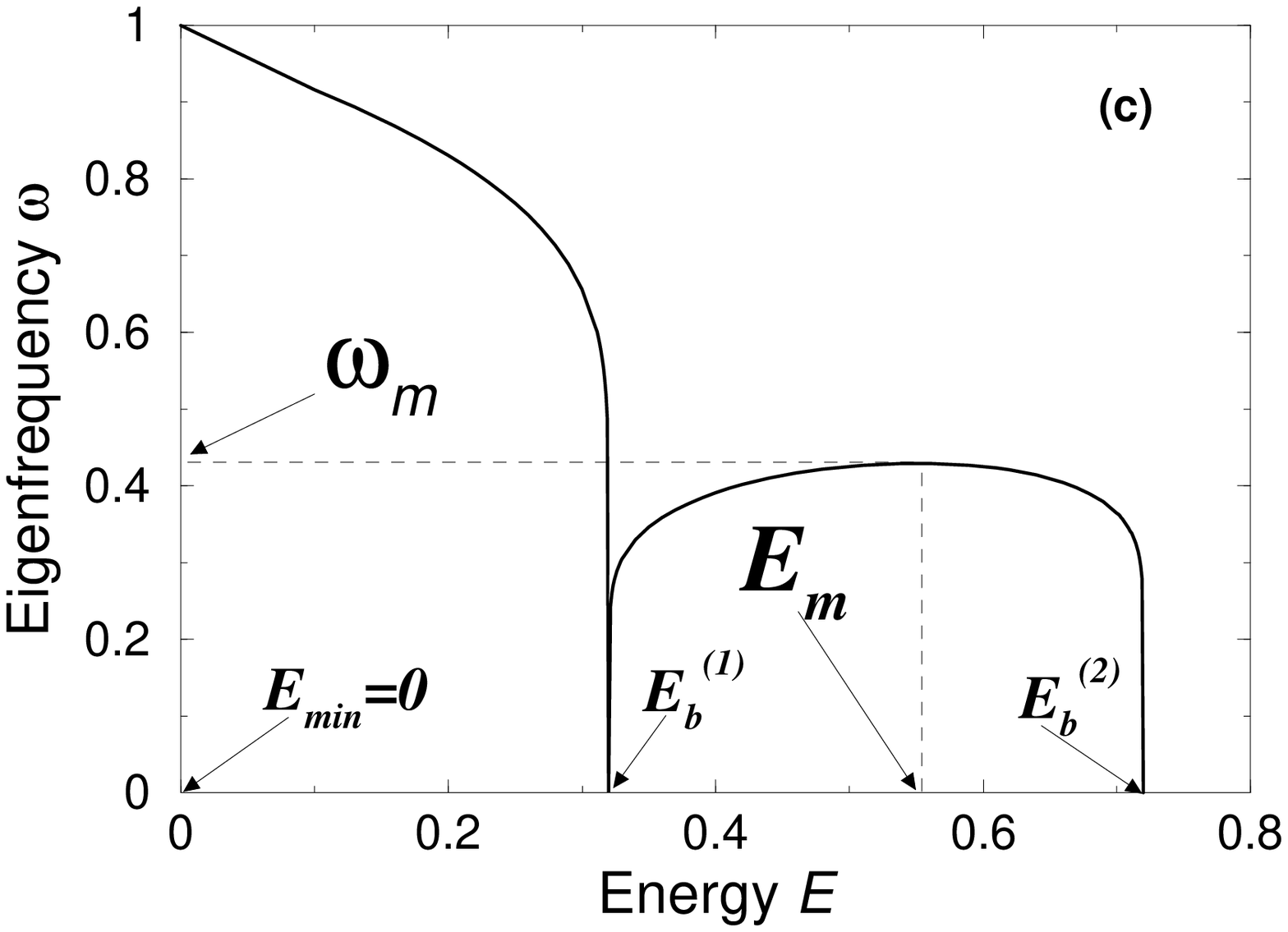,angle=0,width=6.25cm}
\end{picture}
\end{center}
\caption
{The potential $U(q)$, the separatrices in the phase space and
the eigenfrequency $\omega(E)$ for the unperturbed system (1) with $\Phi=0.2$, in
(a), (b) and (c) respectively.
}
\end{figure}

\newpage

\begin{figure}
\centering
{\leavevmode\epsfxsize=2.8 in\epsfysize=2.5 in\epsfbox{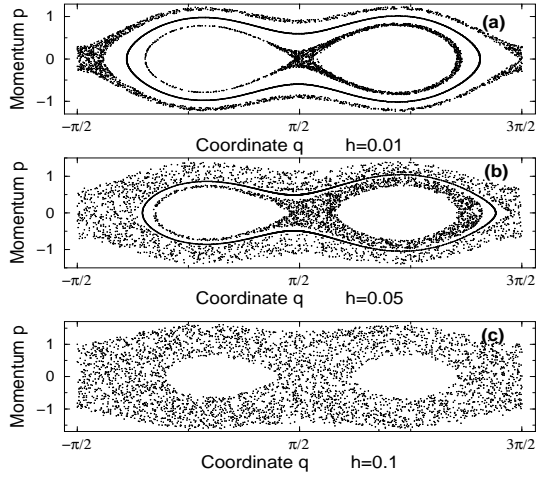}}
\vspace{0.5cm}
\caption{The evolution of the stroboscopic (for instants $n2\pi/\omega_f$
with $n=0,1,2,...$) Poincar\'{e} sections of the system (2)-(1)
at $\Phi=0.2$,
with $\omega_f=0.3$ while $h$ grows from the top to the bottom:
(a) 0.01, (b) 0.05, (c) 0.1. The number of points in each
trajectory is 2000. In parts (a) and (b), three characteristic
trajectories are shown: the inner trajectory starts from the state
$\{I^{(0)} \} \equiv \{p=0,q=\pi/2 \}$ and is chaotic but
bounded in the space; the outer trajectory starts from $\{O^{(0)}
\} \equiv\{p=0,q=-\pi/2 \}$ and is chaotic and unbounded in
coordinate; the third trajectory represents an example of a
regular (KAM)
trajectory separating the above chaotic ones. In (c), the
chaotic trajectories mix.
}
\end{figure}

\begin{figure}
\centering
{\leavevmode\epsfxsize=3.1 in\epsfysize=3.7 in\epsfbox{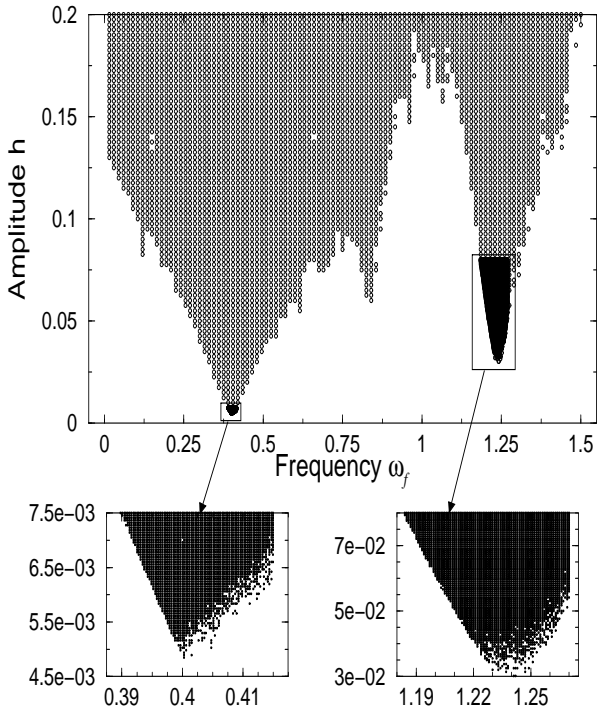}}
\vspace{0.5cm}
\caption{The bifurcation diagram indicating (by shading) the
perturbation
parameters ranges at which global chaos exists. The integration time for each point of the grid is
$12000\pi$.
}
\end{figure}

\begin{figure}
\centering
{\leavevmode\epsfxsize=3.6 in\epsfysize=4.5 in\epsfbox{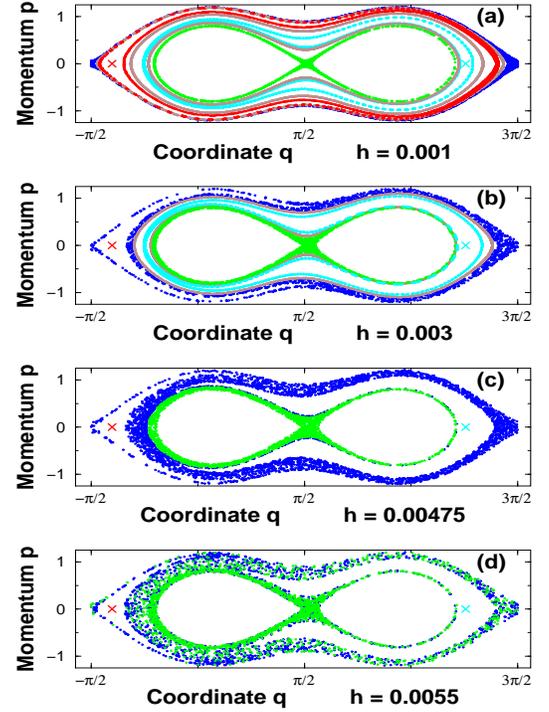}}
\caption{The evolution of the stroboscopic Poincar\'{e} section of the system (2)-(1) with $\Phi=0.2$ as
the amplitude of the perturbation $h$ grows while the frequency is
fixed at $\omega_f=0.401$. The number of points in each trajectory
is 2000. The chaotic trajectories starting from the states
$\{I^{(0)} \} $ and $\{O^{(0)} \} $ are drawn in green and blue
respectively. The stable stationary points (the 1st-order
nonlinear resonances) are indicated by the red and cyan crosses.
The chaotic layers associated with the resonances, in those
cases when they do not merge with those associated with the green/blue chaotic
trajectories, are indicated in red and cyan respectively (their
real width is much less than the width of
the drawing line). Examples of KAM trajectories embracing the
state $\{I^{(0)} \} $ while separating various chaotic
trajectories are shown in brown.
}
\end{figure}


\begin{thebibliography}{99}
\bibitem{more} More complicated cases are considered e.g. in
\cite{pr,Howard:95}.
\bibitem{footnote2}
We still use different colours for
the trajectories starting from $\{I^{(0)}\}$ and $\{O^{(0)}\}$ in order to demonstrate
that their mixing occurs very slowly, which indicates that the
given $h$ is just slightly above the critical value $h_{gc}$.
\end{thebibliography}
\end{document}